\documentclass[runningheads]{llncs}
\usepackage{graphicx}
\usepackage{xcolor}
\usepackage{siunitx}
\usepackage{array}
\newcolumntype{P}[1]{>{\centering\arraybackslash}p{#1}}
\usepackage{cite}
\usepackage{wrapfig}
\usepackage{comment}

\begin{document}
\title{Short Paper: A Centrality Analysis\\ of the Lightning Network}
%
%
\author{Philipp Zabka\inst{1} \and
Klaus-T.\ Foerster\inst{2} \and
Christian Decker\inst{5} \and
Stefan Schmid\inst{3}\inst{1}\inst{4}}

%
%
\institute{Faculty of Computer Science, University of Vienna, Austria \and
Technical University of Dortmund, Germany \and
Faculty of Computer Science, Technical University of Berlin, Germany \and
Fraunhofer SIT, Germany \and
Blockstream, Zurich, Switzerland}
%

\authorrunning{P. Zabka et al.}


\maketitle              
%
\begin{abstract}
Payment channel networks (PCNs) such as the Lightning Network offer an appealing solution to the scalability problem faced by many cryptocurrencies operating on a blockchain such as Bitcoin. 
However, PCNs also inherit the stringent dependability requirements of blockchain. In particular, in order to mitigate  liquidity bottlenecks as well as on-path attacks, it is important that payment channel networks maintain a high degree of decentralization. Motivated by this requirement, we conduct an empirical centrality analysis of the popular Lightning Network, and in particular, the betweenness centrality distribution of the routing system. Based on our extensive data set (using several millions of channel update messages), we implemented a TimeMachine tool which enables us to study the network evolution over time. 
We find that although the network is generally fairly decentralized, 
a small number of nodes can attract a significant fraction of the transactions, introducing skew. 
Furthermore, our analysis suggests that over the last two years, the 
centrality has increased significantly, e.g., the inequality (measured by the Gini index) has increased by more than 10\%.
\end{abstract}

\section{Introduction}\label{sec:intro} 

Blockchain, the technology which is currently revamping the financial sector and which 
underlies cryptocurrencies such as Bitcoin and Ethereum, enables 
mistrusting entities to  cooperate without involving a trusted third party. 
However, with their quickly growing popularity, blockchain networks 
face a scalability problem, and the requirement of performing repeated global consensus
protocol is known to limit the achievable transactions rate.

Payment channel networks (PCNs) are a promising solution to mitigate 
the scalability issue, by allowing users to perform transactions \emph{off-chain}. 
In particular, in a PCN, two users can establish so-called payment channels
among each other, in a peer-to-peer fashion.
The set of channels can then be seen as a graph, 
in which users are represented as nodes and 
channels are represented as edges.
Payments can then also be routed
in a multi-hop manner across these channels
(typically using source routing), with forwarding users typically
charging a small fee. 
Nodes can discover the cheapest routes using a gossip mechanism.
The scalability benefit comes from the fact that it
is only when a channel is opened or closed, that 
changes have to be made to the blockchain. 

By the nature of the service they provide, PCNs need to meet stringent dependability
requirements. 
Interestingly, while over the last years, several interesting approaches to design and 
operate payment channel networks in an efficient and reliable manner 
have been proposed in the literature, relatively little is known 
about the properties of the actually deployed networks today. 

We in this paper are particularly interested in the level of decentralization provided
by PCNs: decentralization is generally one of the key features of blockchain, 
and also naturally required from off-chain solutions. 

Indeed, it has recently been shown that skews in the routing system
(e.g., due to exploits of the payment mechanism), can significantly harm
the network performance, by depleting channels \cite{khalil2017revive}, or even lead to denial-of-service attacks~\cite{aft20} and privacy~\cite{icissp20,tang2020privacy}
and other security issues~\cite{malavolta2019anonymous}.
In order to gain a detailed understanding of Lightning, the most popular PCN, 
we monitored the network for several years, collecting millions of
channel update and gossiping messages. 
To shed light on the network evolution, we further implemented tools which allow us to 
reconstruct the network at previous time stamps.
In this paper, we present the main results of our study of the Lightning Network.
\subsection{Our Contributions}\label{sec:contrib} 

\begin{wrapfigure}{R}{0.45\textwidth}
\vspace{-9mm}
\includegraphics[width=0.45\textwidth]{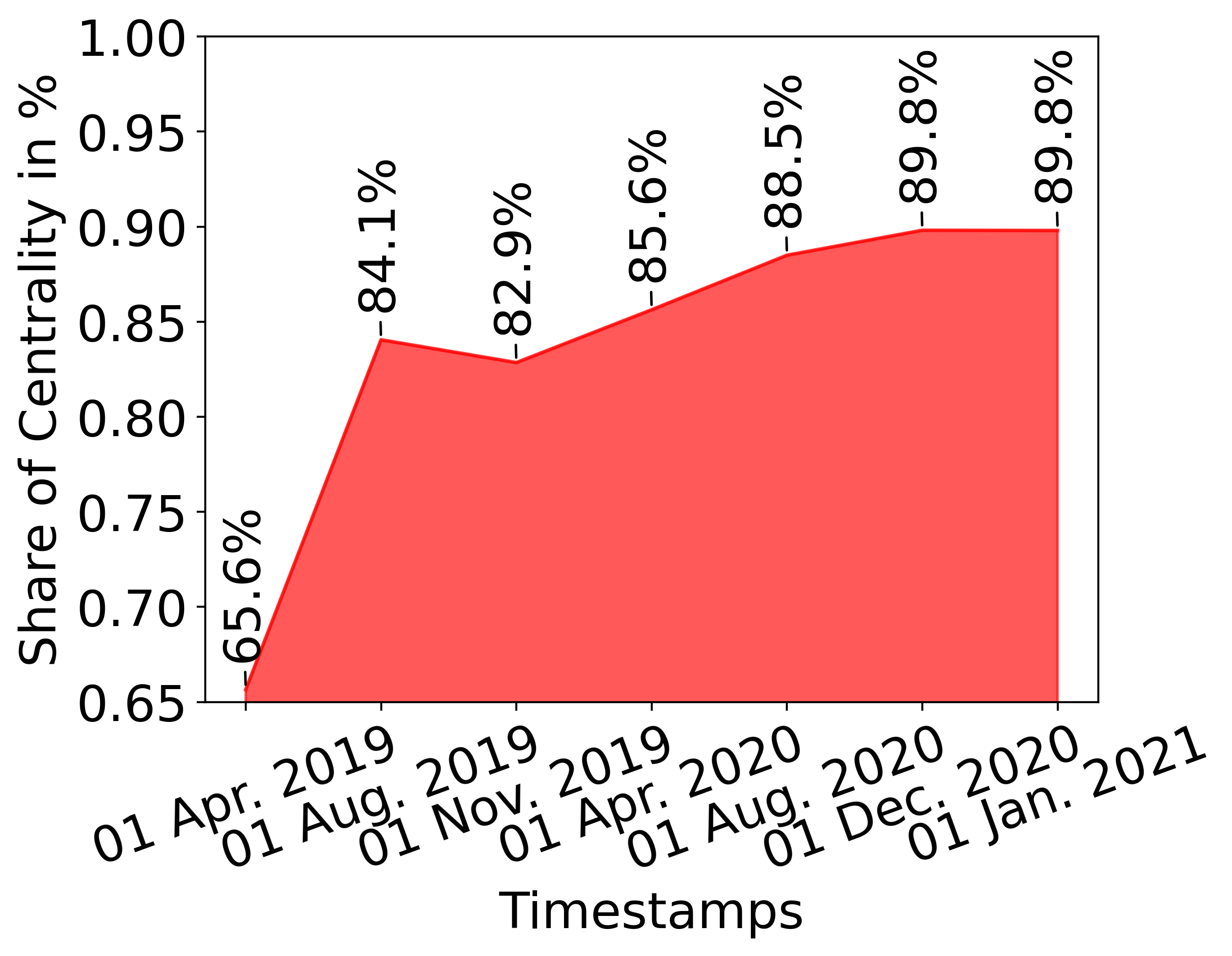}\label{fig:top}
\vspace{-9mm}
\caption{Top 10\% control over routes}
\label{fig:top}
\vspace{-10mm}
\end{wrapfigure}

Motivated by the increasing popularity of payment channel networks
and the resulting performance and dependability requirements,
we report on an extensive empirical study of the most popular
PCN, Lightning. 
In particular, we study to which extent Lightning fulfills the premise
of decentralized transaction~routing.

We find that there is a trend of increasing centralization and a high level of inequality, where 
a small portion of the nodes participate on most transaction routes.
We show that the level of centrality also depends on the transaction size, and we take a look at some of the highest ranked nodes according to centrality. We uncover that a fair share of nodes remained at the top over the examined period.
To just give one example, our analysis shows that 
the top 10\% of all nodes control a vast majority of all transaction routes, 
and that the controlled share increases over time, 
see Fig.~\ref{fig:top}.

For our study, we collected significant data from the live Lightning Network, over a time span of almost two years.
This data includes over 400k node announcement
messages, over 1m channel announcement messages, and over 6m channel update messages.
We further developed \emph{TimeMachine}, 
a tool which allows us to reconstruct the network at desired moments in time. 
We accomplish this with the help of the above mentioned gossip mechanism. 

As a contribution to the research community, in order to ensure reproducibility as well as to support future research in this area,
we make available all our code and experimental artifacts \cite{artifacts} together with this paper.

\subsection{Related Work}\label{sec:relwork} 

Over the last years, many interesting approaches to design and operate payment channel networks have been proposed in the literature, 
often accounting for dependability aspects \cite{kapposEmpiricalAnalysisPrivacy2021,Roher2020Counting,romitiCrossLayerDeanonymizationMethods2021,Harris2020Flood,moreno-sanchezPrivacyPreservingPayments2015,malavoltaAnonymousMultiHopLocks2019}, 
and we refer the reader to \cite{dotan2021survey,gudgeon2020sok,neudecker2018network} for an overview.

In this paper, we are particularly interested in issues related to centralization,
a topic which has recently also received much attention in the context of Bitcoin in general \cite{coindesk,beikverdi2015trend,forbes}. 
In the context of PCNs, it has been shown that centralization of the routing system can harm 
performance~\cite{aft20,spam}, liquidity~\cite{khalil2017revive,khamis2021demand}, security~\cite{malavolta2019anonymous}, 
and privacy aspects~\cite{tang2020privacy,icissp20,malavolta2017concurrency,tripathy2020mappcn}, especially when considering
on-path adversaries. 

Interestingly, relatively little is known about the empirical properties of deployed payment channel networks.
The Lightning Network's topology has been analyzed by Seres et al.~\cite{seres2020topological}.
Their work studies the robustness of the network against random failures of nodes as well as attacks targeting nodes.
A similar, but more in detail work has been carried out by Rohrer et al.~\cite{RohrerMT19}.
Martinazzi et al.~\cite{martinazzi_evolving} analyzed the evolution of the Lightning Network over a period of one year, beginning on its launch on the Bitcoin mainnet in January 2018. 
Their work focuses on the topological robustness of the network, e.g., against attacks, where they also detect a high influence of a few nodes on the network. 
Next, a large scale empirical analysis on the client and geographical classification of nodes is performed by Zabka et al.~\cite{zabkaEmpAnalysis}, see also Mizrahi et al.~\cite{DBLP:ln_congestion_mizrahi}. 
Related to this, Scellato et al.~\cite{ScellatoMML10} study how geographic distance affects social ties in a social network and Mislove et al.~\cite{MisloveLAOR11} examine geographical, gender and racial aspects of Twitter users to the U.S. population.

\subsection{Organization}\label{sec:organization} 
\noindent\textbf{Organization.} The remainder of this paper is organized as follows. Section \ref{sec:prelim} introduces some preliminaries and Section \ref{sec:method} describes our methodology, followed by the centrality analysis in Section \ref{sec:centrality}.
We subsequently conclude in Section \ref{sec:conclusion}.

\section{Preliminaries}\label{sec:prelim} 

We now introduce some of the necessary basics of the Lightning Networks and some specific preliminaries for the remainder of the paper. 

\smallskip
\noindent\textbf{The Lightning Network.}
The Lightning Network is an off-chain solution to improve the scalability of cryptocurrencies such as Bitcoin. The network can be accessed via three clients, namely LND \cite{lndgit} implemented in Go, C-Lightning \cite{clightninggit} implemented in C and Eclair \cite{eclairgit} implemented in Scala. However, with an usage of more than 85\%, LND is currently by far the most popular client \cite{zabkaEmpAnalysis}. 
The Lightning Network users are able to create bidirectional connections to other users, called channels. These channels can be used to send instant payments between two users, which do not need to be necessarily directly connected. If a payment is routed across multiple users, the users in between the route may demand fees for the routing process.  
The Lightning Network does not operate on the blockchain itself, however the first transaction called the funding transaction to create a channel needs to be propagated onto the blockchain. The same goes for the last transaction or closing transaction to end the connection between two users. All intermediary transactions are not propagated onto the blockchain and therefore can be processed in a much faster fashion.

\noindent\textbf{Gossip Messages.}
As the name implies, gossip messages are propagated through the whole network to either announce a node or channel creation or an update. Therefore, all participants have an contemporary view of the network. This mechanism is especially important in the case that a node wants to route a payment to a node it is not directly connected with. In the following we will take a more in detail look
at the three most important gossip messages, which are specified in the Basics of the Lightning Technology (BOLT) \cite{bolt7}:

\begin{itemize}
  \item \textbf{node\_announcement\_message}: This message allows nodes to inform other participants about extra data associated with it, besides the node ID. It contains data such as the IP address, color, alias and timestamp as well as information for opting into higher level protocols.
  \item \textbf{channel\_announcement\_message}: If a channel is created between two nodes this message is propagated through the network. It contains information such as an short channel ID, which is an unique identifier for the channel, as well as both node IDs. 
  \item \textbf{channel\_update\_message}: A channel is practically not usable until both sides announce their channel parameters. These parameters are announced in this message. As the Lightning Network is directed, both channel participants have to send a message. The parameters included in this message are among other things used to calculate the routing fees. Every time one side updates its channel parameters, this message is broadcast in the~network.
\end{itemize}

\noindent\textbf{Routing Fees.}\label{routing_fees}
In the Lightning Network nodes along a routed path take a small fee for forwarding transactions.
The parameters necessary for the calculation are \textit{fee\_base\_msat} and \textit{fee\_proportional\_millionths} which can be found in the \textit{channel\_update\_message}. Hereby \textit{fee\_base\_msat} denotes the constant fee a node will charge for a transfer and \textit{fee\_proportional\_millionths} is the amount a node will charge for each transferred satoshi over their channel. Fees are calculated as follows, where {transferred\_amount} denotes the transaction in millisatoshi:

\vspace{2mm}
 fee\_base\_msat$+$(transferred\_amount$*$fee\_proportional\_millionths$/$\num{1000000})


\vspace{2mm}

{
\noindent\textbf{Betweenness Centrality.} 
The betweenness centrality represents a measure in a network based on shortest paths, a node's centrality is based on how many such paths traverse it.
Formally, the betweenness centrality $c_{B}$ of the nodes $v\in V$ is $c_{B}(v) = \sum_{s,t \in V}{\sigma(s,t|v)}/{\sigma(s,t)}$, with $\sigma(s,t)$  [$\sigma(s,t|v)$] as \# shortest $st$-paths [through $v$, $v \neq s,t$]. If $s=t$, $\sigma(s,t)=1$, and if $v\in s,t$, $\sigma(s,t|v)=0$~\cite{SciPyProceedings_11,DBLP:journals/socnet/Brandes08}.
For every node pair in a connected unweighted graph, there exists at least one shortest path between these nodes such that the number of edges is minimized. 
For weighted graphs such as the Lightning Network, where channel routing fees represent edge weights, the sum of the edge weights is minimized. 

Among several interesting alternatives~\cite{liu2011Controllability, DasSP18}, we focus on betweenness centrality as our main centrality measurement. 
Nodes with high betweenness centrality  have a considerable amount of influence on a network by means of information control, since most of the network traffic will pass though them---in contrast to other centrality measures which represent a more local view, e.g., degree centrality, which counts the numbers of edges incident to a node.

A high betweenness centrality is a particular concern as nodes choose routing paths with the overall cheapest fees, and a skewed centrality indicates that routing paths are concentrated to a small subset of nodes. 
A skewed centrality may not only quickly deplete payment channels, but also makes the network vulnerable: many attacks recently reported in the literature are based on on-path adversaries \cite{RohrerMT19, DBLP:ln_congestion_mizrahi}. Getting a significant amount of traffic can also raise privacy concerns, e.g., during route discovery.
}


\section{Methodology}\label{sec:method} 

We next introduce the methods to obtain and process our data set.

\noindent\textbf{TimeMachine.}
The Lightning Network TimeMachine \cite{lngossip-unblind} is a tool written in Python, which  reconstructs the state at a prior point in time by replaying gathered gossip messages up to that point in time. 
We have deployed a number of C-Lightning nodes that collect and archive these messages, which are then deduplicated and ordered by their timestamp, in order to allow the TimeMachine to replay them in the correct order, and terminate once the desired point in time has been reached, leaving the view of the network close to what the public network would have looked like at that time. 
We utilized the TimeMachine to rebuild the network at seven different points in time, covering a time span of two years ranging from 01 Apr. 2019 to 01 Jan. 2021. We then used the Python library NetworkX \cite{SciPyProceedings_11} to further analyze the networks in regard to the betweenness distribution in different timestamps. 
With the help of our TimeMachine we were able to reconstruct the network as it was at the timestamps mentioned in Table \ref{tab_timestamps}. From now on we will reference the timestamps as T1 - T7. 

\begin{wraptable}{r}{6.2cm}
\vspace{-1.3cm}
\caption{Lightning Network Snapshots}\label{tab_timestamps}
\begin{tabular}{cccc}
\hline
Abbr.\ &  Timestamp & Date & \# Nodes\\
\hline
 {\bfseries T1} &  1554112800 & 01 Apr. 2019 & 1362\\
 {\bfseries T2} &  1564653600 & 01 Aug. 2019 & 4589\\
 {\bfseries T3} &  1572606000 & 01 Nov. 2019 & 4699\\
 {\bfseries T4} &  1585735200 & 01 Apr. 2020 & 5230\\
 {\bfseries T5} &  1596276000 & 01 Aug. 2020 & 5905\\
 {\bfseries T6} &  1606820400 & 01 Dec. 2020 & 6331\\
 {\bfseries T7} &  1609498800 & 01 Jan. 2021 & 6629\\
\hline
\end{tabular}
\vspace{-0.8cm}
\end{wraptable}

\vspace{1mm}
\noindent\textbf{Data Set.}
Our data was collected with help of C-Lightning nodes, which synchronize their view of the network topology by listening and exchanging gossip messages. Internally C-Lightning will deduplicate messages, discard outdated \textit{node\_announcements} and \textit{channel- \_updates}, and then apply them to the internal view. In order to persist the view across restarts, the node also writes the raw messages, along some internal messages, to a file called the \textit{gossip\_store}. The node compacts the \textit{gossip\_store} file from time to time in order to limit its growth. Compaction consists of rewriting the file, skipping messages that have been superceded in the meantime.
%
Our data set is comprised of the three gossip messages discussed in the previous section. Our nodes have recorded more than  \num{400000} \textit{node\_announcement messages}, more than \num{1000000} \textit{channel\_announcement messages}, and over \num{6400000} million \textit{channel\_update messages}. 

\section{Centrality Analysis}\label{sec:centrality} 

This section reports our main results from the centrality analysis. We performed a detailed analysis where we measured the betweenness centrality, a major centrality measure, of the Lightning Network at different points in time and observed how it has developed over almost two years. More precisely, we took seven snapshots of the network, dating from  01 Apr. 2019 to 01 Jan. 2021. Based on the formula for calculating routing fees introduced in Section \ref{routing_fees} we calculated the betweenness of each node based on three different realistic transaction sizes namely \num{10000000} Millisatoshi (0.0001 BTC), \num{1000000000}  Millisatoshi (0.01 BTC) and \num{10000000000} Millisatoshi (0.1 BTC).
The idea of calculating the betweenness with different transaction sizes was if we could detect significant changes.

\subsection{Historic Betweenness Analysis of the Lightning Network}\label{sec:historic} 

Evaluating the Lightning Network at different points in time in terms of the betweenness centrality can provide us with insights which allow us to better comprehend how it has developed until now e.g. has it become more centralized or the opposite and also make predictions in which direction it may develop in the future. We start by we examining our latest snapshot first. 

\smallskip
\noindent\textbf{Timestamp T7}
We decided to use a logarithmic scale on the x-axis to better display the long range of centrality values (1 - \num{7500000}). Further, we do not include nodes with a centrality value of 0, as they merely represent leafs in the graph. Also the amount of leaf nodes is astonishing high, up to \num{5520} nodes out of \num{6630} in T7, and would distort the graph.

\begin{figure*}[b]
\vspace{-6mm}
      \includegraphics[width=0.32\textwidth]{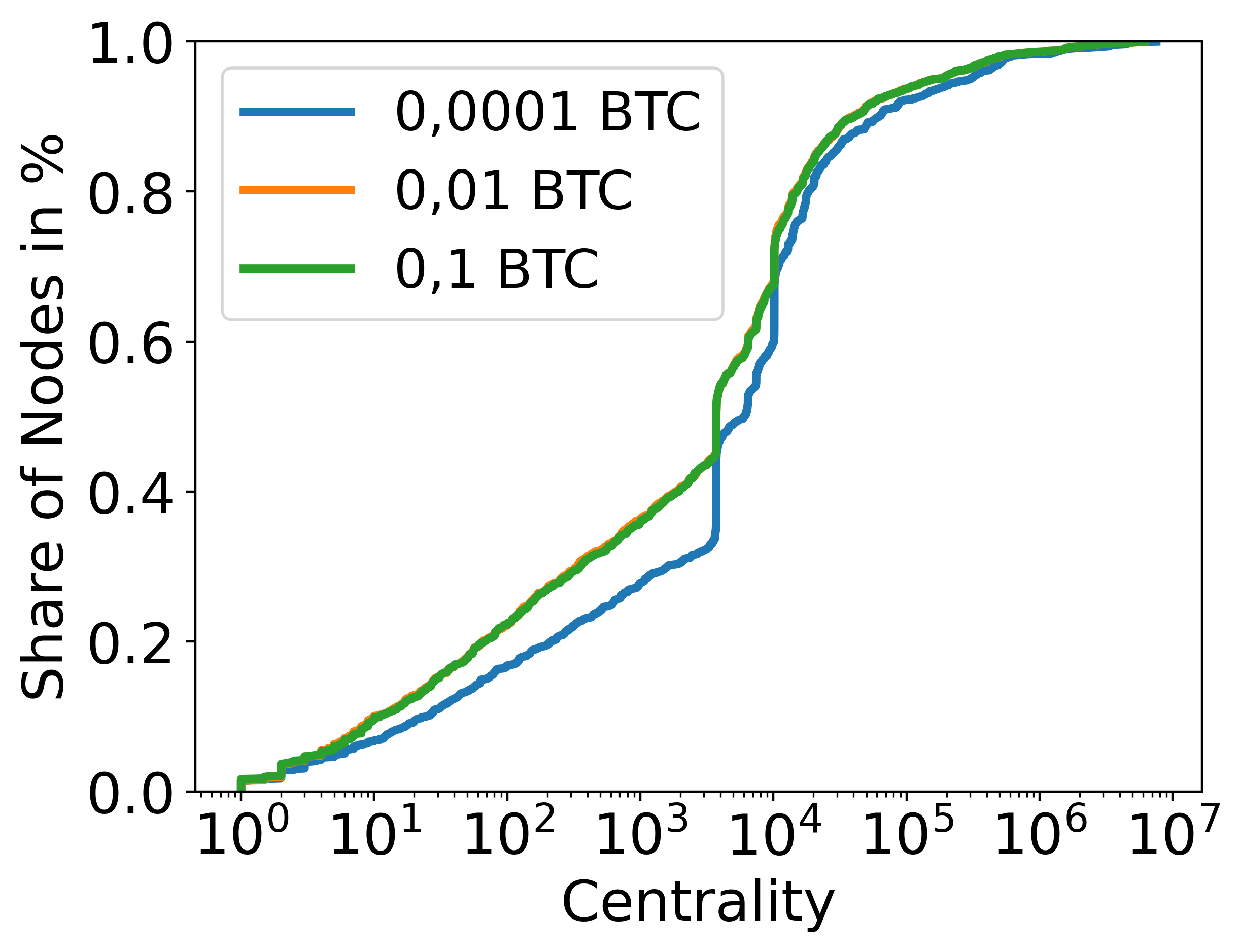}
      \includegraphics[width=0.32\textwidth]{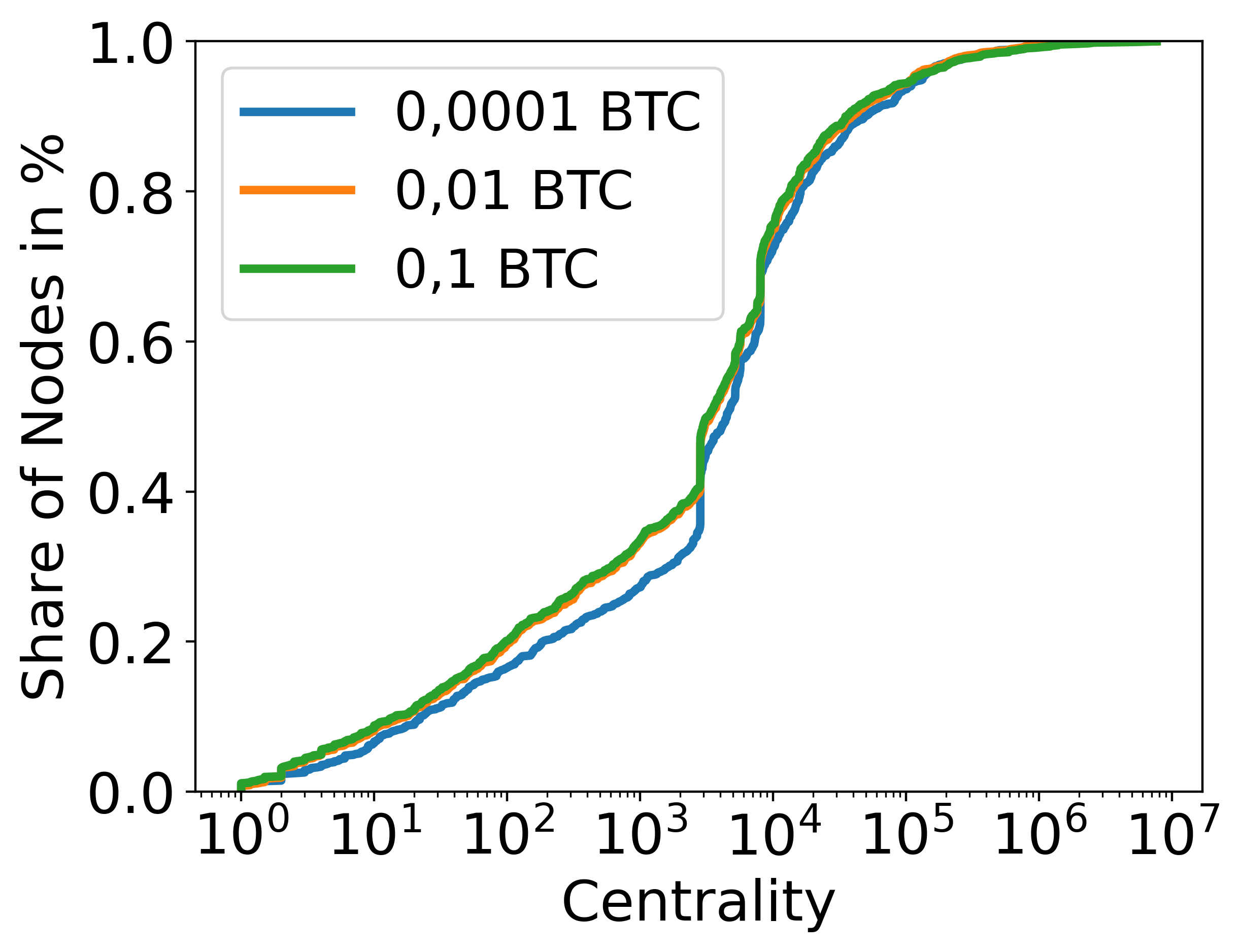}
      \includegraphics[width=0.32\textwidth]{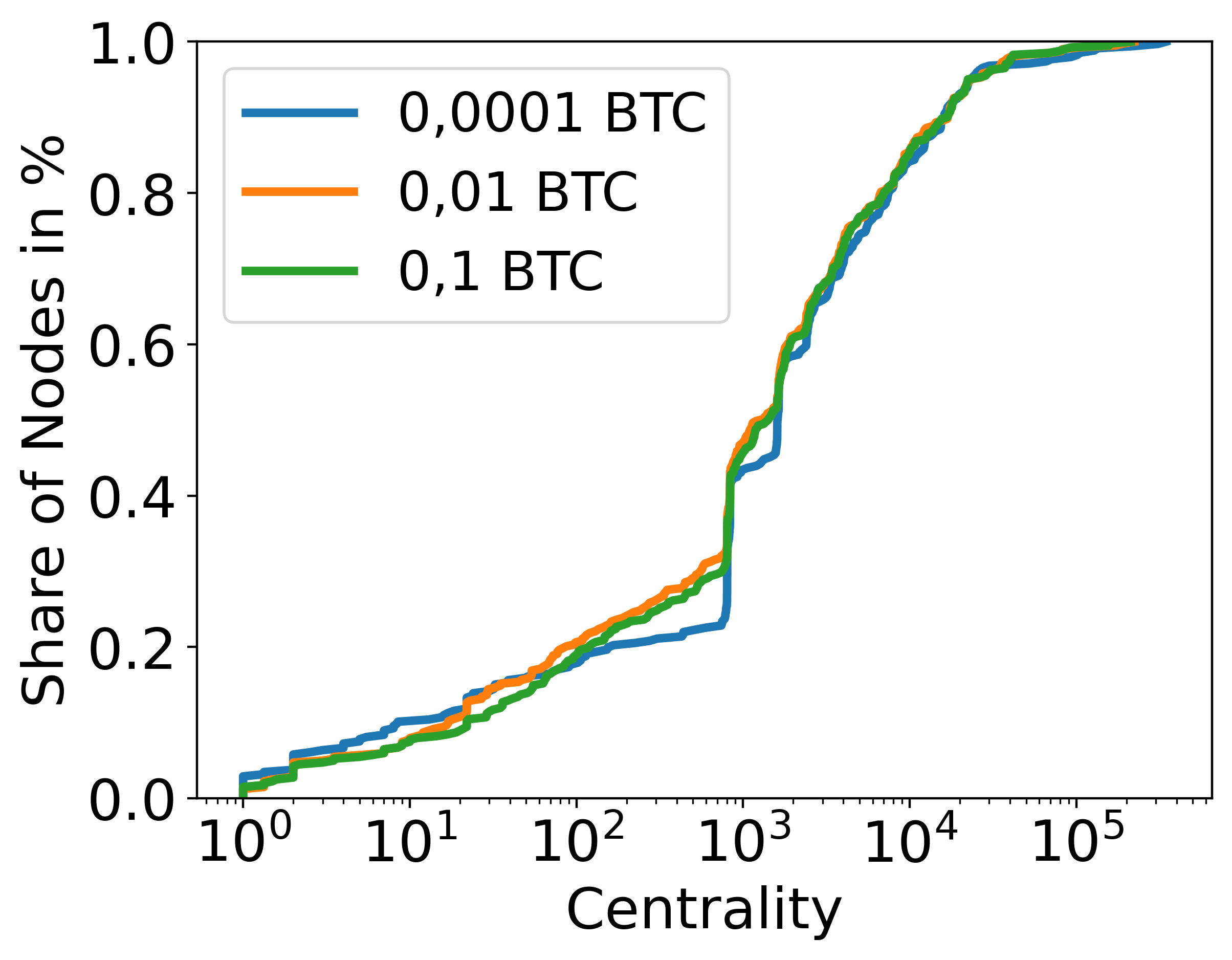}
      \vspace{-4mm}
      \caption{Centrality distribution in timestamps T7 (left), T4 (middle) and T1 (right)}
      \label{fig:centrality_distribution}
\end{figure*}

In Fig. \ref{fig:centrality_distribution} (left) we can see that transaction size has indeed an influence on a node's centrality if the transaction amount is low or high enough. In the case of 0.1 BTC respectively 0.01 BTC there is almost no change in the centrality distribution among the nodes, however, in the case of 0.0001 BTC we can see a significant shift. A possible explanation for this shift in distribution we are experiencing is that for smaller transactions, different routes are calculated. The next noticeable observation is the high jump around the 4000 betweenness centrality mark for all three transaction sizes. For 0.0001 BTC roughly 100 nodes are affected and for 0.01 BTC or respectively 0.1 BTC roughly 80 nodes are concerned. A more in-depth analysis would be required to fully comprehend this phenomenon, but a possible cause can be that these nodes are all positioned on a specific shortest path and therefore share the same centrality.

Another interesting observation is that although the centrality of the majority of nodes is lower when calculated with the lowest transaction size, the centrality of the most central node is the highest of all three transaction sizes with \num{7500000}. For comparison the centrality for 0.1 BTC and 0.01 BTC caps at \num{6100000}.

\smallskip
\noindent\textbf{Timestamp T4}
In T4 we can make out only a few detailed changes 9 months prior to our latest timestamp T7. Observing Fig. \ref{fig:centrality_distribution} (middle) shows the centrality distribution for 1026 nodes out of 5231, so 4205 nodes remain leaf nodes with a centrality of 0. We can detect a similar jump at a centrality of approximately 3000 with 65 nodes having the exact same score. Another jump occurs at the 8000 mark with 48 nodes having the same value.

As was already the case in T7, the higher the centrality gets the more closer the share of nodes is that has a similar high centrality. However, this is due to the fact that only a few nodes share such a high betweenness centrality.

\smallskip
\noindent\textbf{Timestamp T1}
Fig. \ref{fig:centrality_distribution} (right) depicts the centrality distribution for T1, which is 21 months prior to T7. At the first glance we can immediately detect that now all transaction sizes have a much more similar impact on the centrality distribution of the nodes in the network. However, this is most probably due to the overall lower amount of nodes in the network at that point in time and therefore limited amounts of paths that can be selected. According to our data, there are 1361 nodes in the network in T1 and only 347 out of them have a higher centrality than 0. 

The graphs are rather similar, but jumps still occur. Betweenness values calculated with the transaction size of 0.0001 BTC experience the highest jumps. The first one starts at around 1000 and affects 0.3\% of the nodes, the second one starts at around 1600 and affects 0.2\% of the nodes. At last, compared to the most central node in T7, the most central node in T1 only reaches an centrality of \num{350000}. Even though the lower value is the result of fewer nodes in the network, one can not deny the rapid centralization of the network within the period of two years. We next further substantiate our observation of growing centrality. 

\subsection{Inequality in the Lightning Network}\label{sec:gini}

\begin{figure*}[b]
\vspace{-4mm}
      \includegraphics[width=0.32\textwidth]{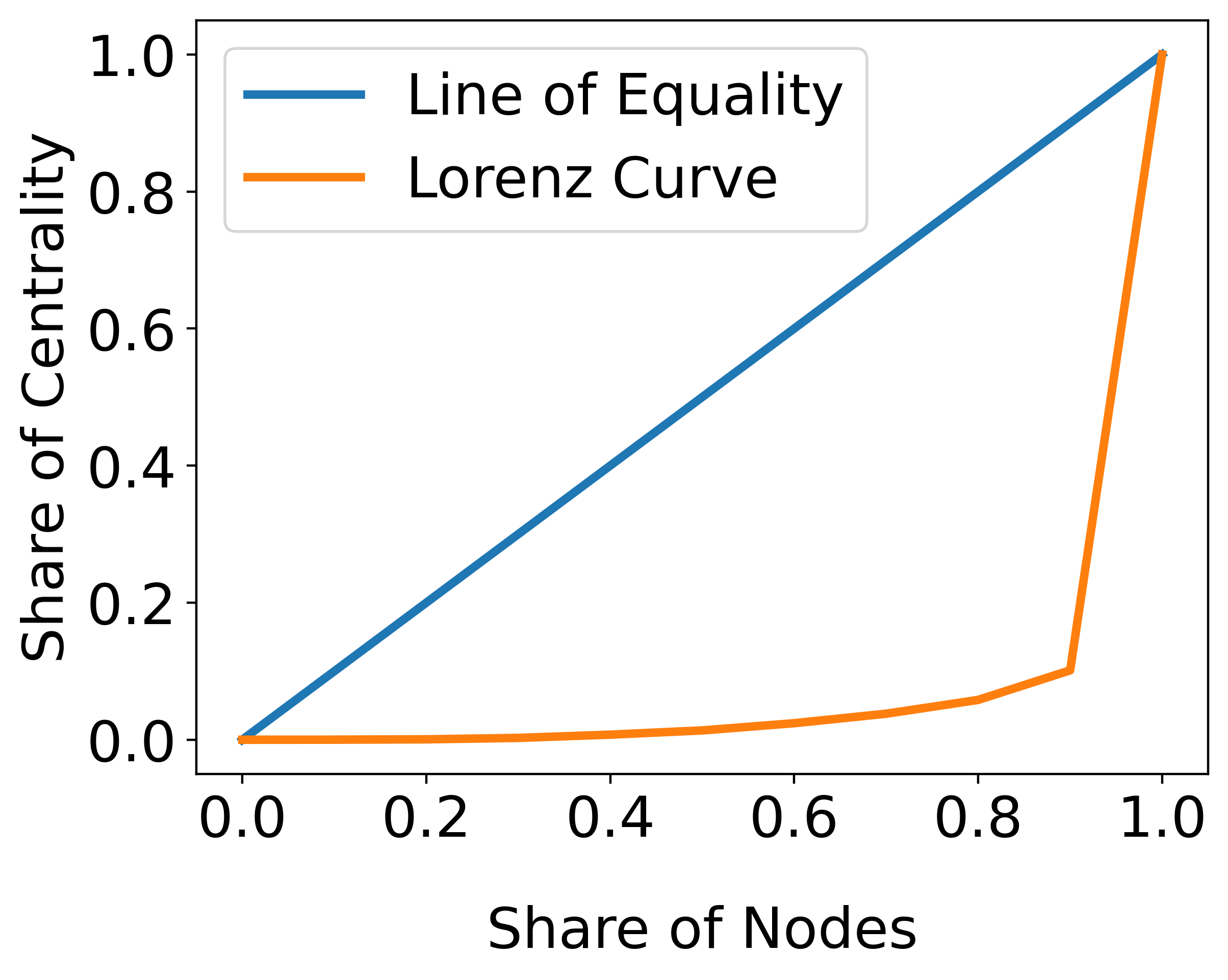}
      \includegraphics[width=0.32\textwidth]{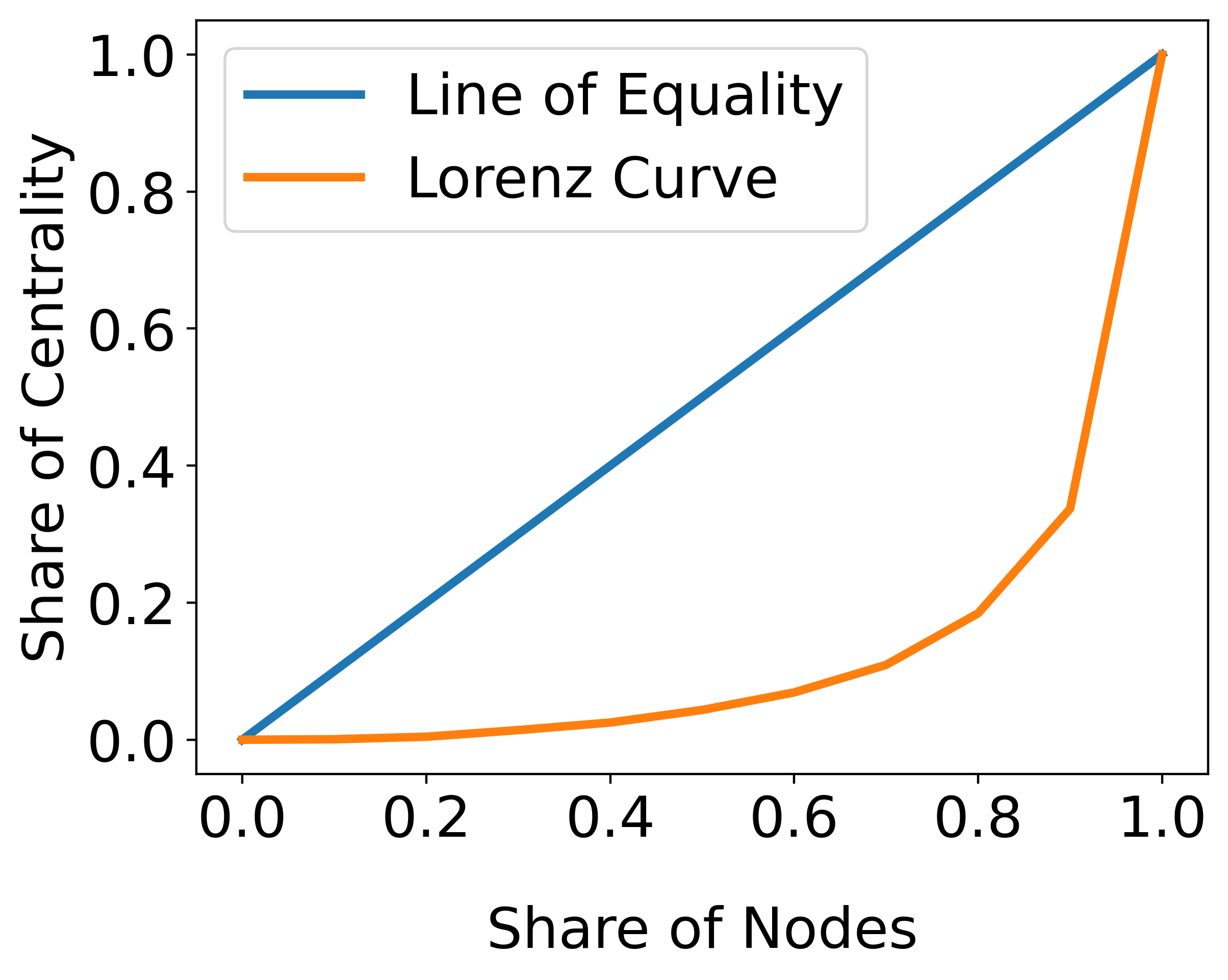}
      \includegraphics[width=0.32\textwidth]{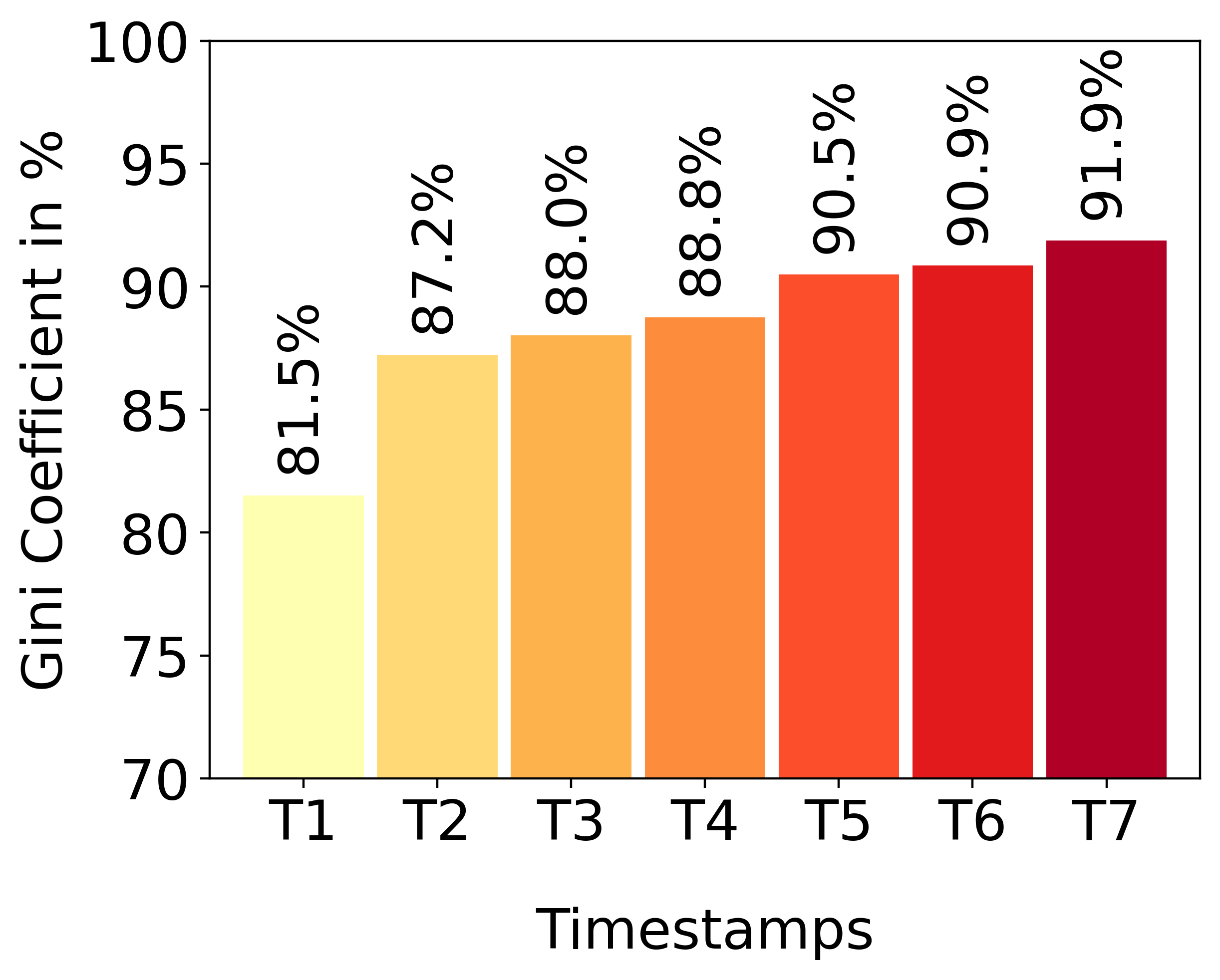}
  
      \vspace{-4mm}
      \caption{Lorenz curves for the timestamps T7 (left) and T1 (middle). Gini Coefficients ranked according to all seven timestamps (right)}
      \label{fig:lorenz_curves}
\end{figure*}
{

The Gini coefficient is an economic measure for the inequality within a nation or a social group. Similarly, we use this index in the context of payment channel networks to shed light on the inequality and skew there exists in the network topology. In particular, an "unfair" distribution concentrates much control to a small set of nodes, which is problematic not only for the efficiency of the network but also raises security concerns. Many attacks in the literature are based on on-path adversaries \cite{RohrerMT19, DBLP:ln_congestion_mizrahi}, which hence have significant control. This also generally goes against the idea of decentralization of finance.
}


Figure \ref{fig:lorenz_curves} (left and middle) depicts the Lorenz curves for T7 and T1. The Gini coefficient is equal to the area below the line of perfect equality minus the area below the Lorenz curve, divided by the area below the line of perfect equality.
Looking at Fig. \ref{fig:lorenz_curves} (left) showing the latest snapshot of the network, we can see an excellent example of a perfectly unequal distribution, where 90\% of the nodes only correspond to 10\% of the cumulative betweenness of all nodes. Consequently, this indicates an extraordinarily high network centralization, where 90\% of the shortest paths in the network lead through only a few highly centralized nodes. 
Next, looking at Fig. \ref{fig:lorenz_curves} (middle) we can observe that 90\% of the nodes make up for slightly more than 30\% of the betweenness, which is still not an ideal scenario. Subsequently, we can conclude from our observations that within 21 months the centralization has risen by 20\%.
Fig. \ref{fig:lorenz_curves} (right) depicts the Gini coefficients for all seven timestamps. Here we observe an upward trend in the direction of inequality or centralization. The coefficient is slightly rising each timestamp, with the biggest jump with absolute 5\% being between T1 and T2. Overall, we can deduce that the Lightning Network is highly centralized. Having only few, very influential nodes through which most paths are routed, is not beneficial for the robustness of the network. These nodes pose as significant targets for attacks and could disrupt the network in the case of failure. However not only attackers could exploit this situation, but also the nodes or rather the individuals controlling these nodes. 

\subsection{Analysis of the Top 10 Nodes}
\label{sec:inflnodes} 

\begin{wrapfigure}{R}{0.45\textwidth}
      \centering
      \includegraphics[width=0.32\textwidth]{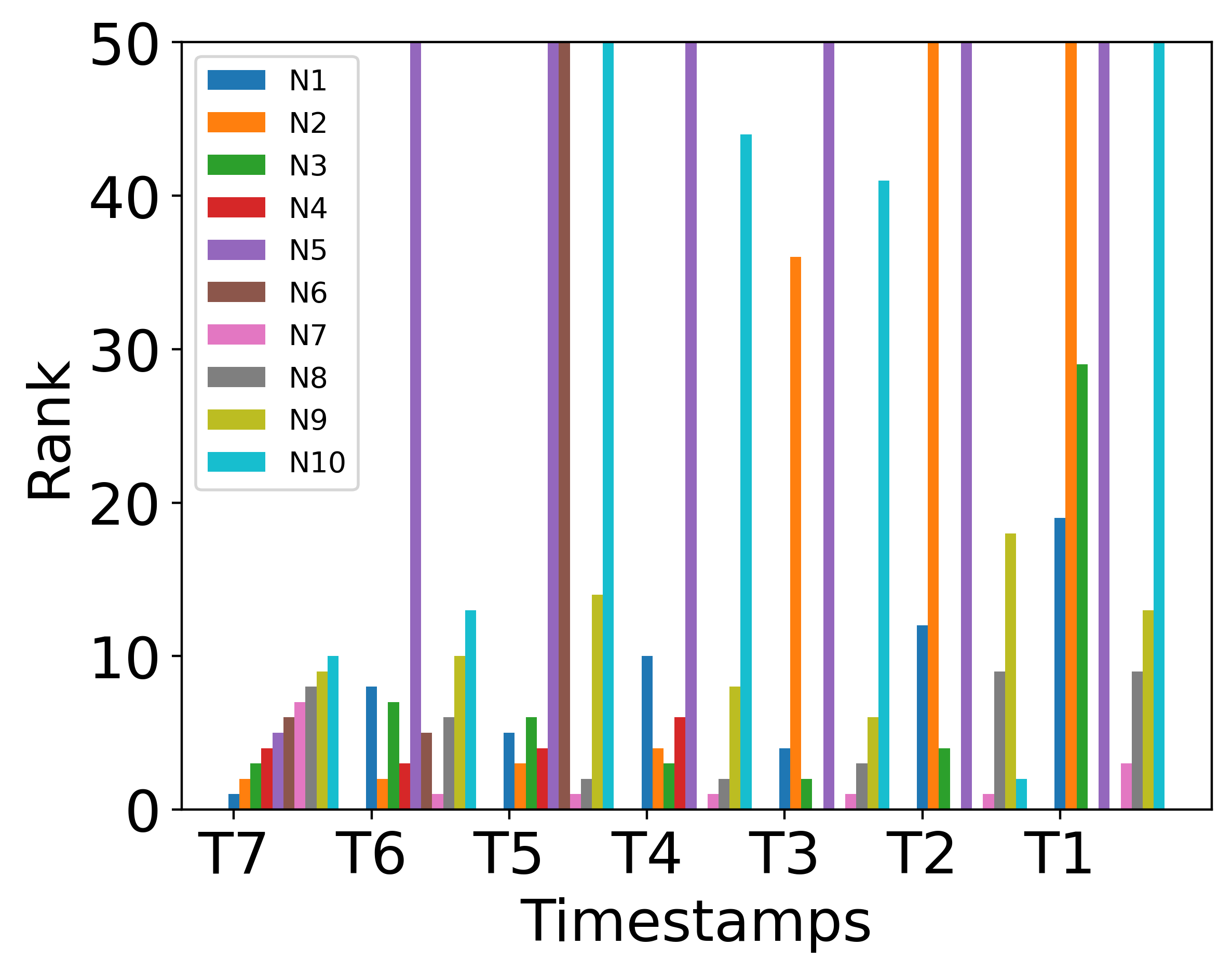}
      \includegraphics[width=0.32\textwidth]{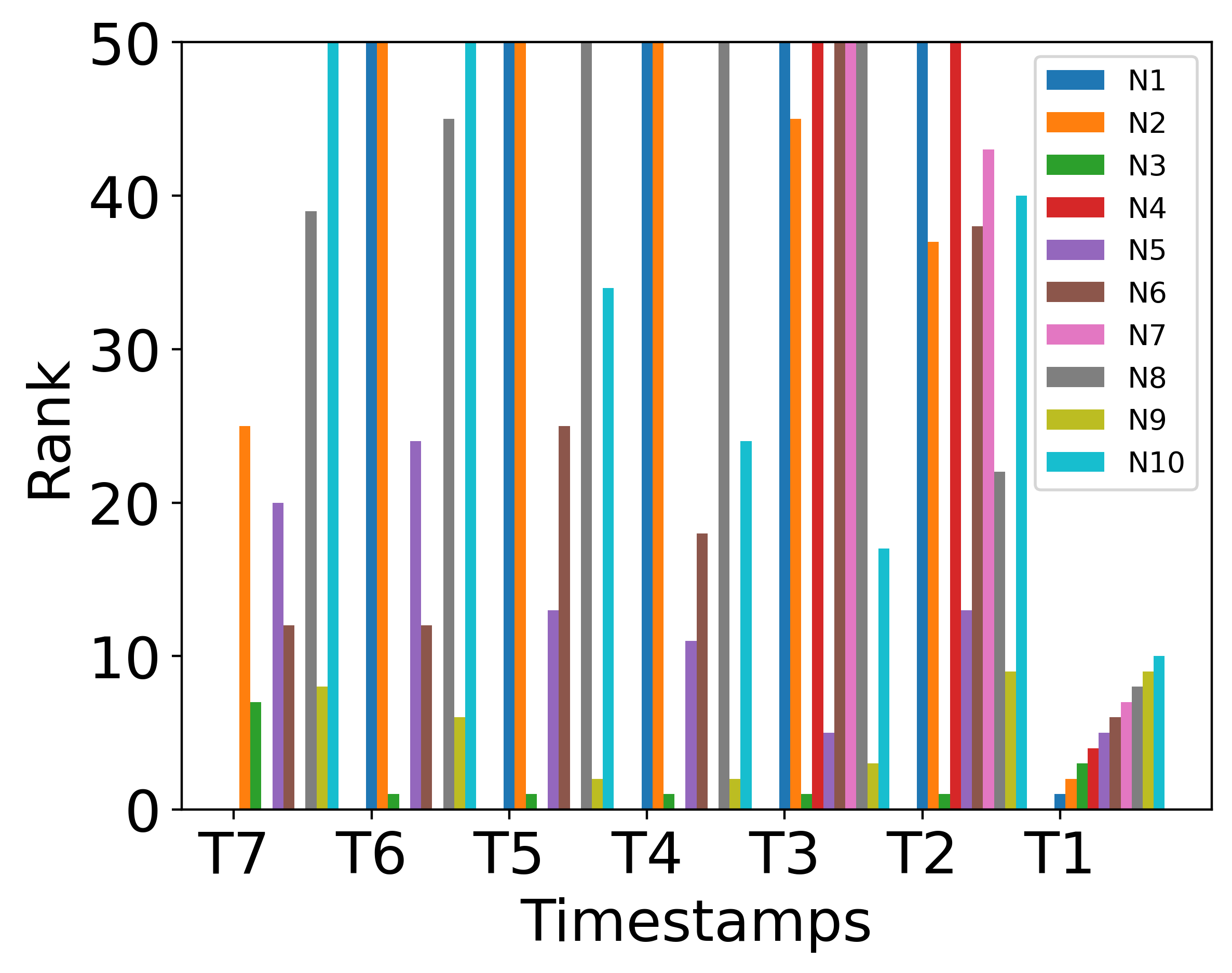}
      \vspace{-4mm}
      \caption{Top ten influential node timelines, with latest left and oldest right}
      \label{fig:top_10}
       \vspace{-8mm}
\end{wrapfigure}

We lastly trace the performance of the most influential nodes, based on their centrality, in our latest and oldest timestamp, and briefly discuss our findings.

Fig.~\ref{fig:top_10} (left) depicts the top 10 nodes with the highest centrality in the latest timestamp T7 and their ranks in the earlier timestamps.
We can see that most top nodes were also highly ranked in the past, e.g., N1 has always been in the Top 20 --- with some nodes starting to appear later, but then already at high rank, such as N3 (ACINQ \cite{acinq}, developer of Eclair).

We now look the other way around to observe if a node could hold its central position in the network. Fig. \ref{fig:top_10} (right) depicts the top 10 nodes in T1 our oldest snapshot and how the nodes performed from there on.
For clarification the nodes depicted in this figure are partially not same as in Fig.\ref{fig:top_10} (left). 
Many nodes could not hold their position, the only nodes which stayed in the Top 10 through all timestamps are N3~\cite{rompert} and N9 or respectively N7 and N8 in Fig. \ref{fig:top_10} (left).

Hence, we see that many powerful nodes of today were already highly influential in the past, respectively came in with a strong backing.
Yet, a strong position in the past is not a guarantee, and many past top 10 nodes lost~influence.

\section{Future Work}\label{sec:conclusion} 


We believe that our work opens several interesting directions for future research.
In particular, it will be interesting to 
investigate other off-chain networks,
further implications of centrality in cryptocurrency networks such as censorship concerns,
and to develop mechanisms to foster more decentralization in payment channel networks.
The latter includes the design of alternative, incentive-compatible routing mechanisms. 

\section*{Acknowledgements and Bibliographical Note}
This paper \cite{zabkaCentralityAnalysis} appears at the Financial Cryptography and Data Security 2022 \cite{fc22} and we thank our shepherd Karim Eldefrawy and the anonymous reviewers of Financial Cryptography and Data Security 2022 for their time and suggestions on how to improve the paper.
This project has received funding from the Austrian Science Fund (FWF) project ReactNet (P 33775-N), 2020-2024.


%
%




\bibliographystyle{FC22}
\bibliography{FC22}

\end{document}